\documentstyle[psfig,twocolumn,prl,aps,amssymb]{revtex}
\begin{document}
\wideabs{
\draft

\date{\today} \title{How universal is the fractional-quantum-Hall edge Luttinger
liquid?}
\author{Sudhansu S. Mandal and J.K. Jain}
\address{Department of Physics, 104 Davey Laboratory, The Pennsylvania State 
University, University Park, Pennsylvania 16802}
\maketitle

\begin{abstract}

This article reports on our microscopic investigations of 
the edge of the fractional quantum Hall state at filling factor $\nu=1/3$.  We 
show that the interaction dependence of the wave function is well 
described in an approximation that includes mixing with higher 
composite-fermion Landau levels in the lowest order. 
We then proceed to calculate the equal time edge Green function, which provides 
evidence that the Luttinger exponent characterizing the decay of the Green function
at long distances is interaction dependent.  The relevance of this result to 
tunneling experiments is discussed.

\end{abstract}

}

There has been interest in the physics of edge states in  
fractional quantum Hall effect (FQHE) \cite{Tsui} for several reasons.  
On the quantum Hall plateaus, 
the Hall current flows along the edges (at least for sufficiently small
currents) because that is the only place where gapless excitations are present.
Presence of extended edge states is crucial for the QHE, as is 
clear in the explanation of the physics of the plateaus \cite{plateau}. 
The physics of the edge states is also directly relevant to  
experiments studying tunneling into edge states \cite{Chang1,Chang2}.

Theoretically, the electrons at the edge of a FQHE system  
constitute an example of a one dimensional liquid, the long-distance, low-energy 
physics of which is generically described by the Luttinger model.
One of the most interesting aspects of the edge Luttinger liquid was 
the assertion, based on general arguments,
that the quantization of the Hall conductance fixes the parameter
characterizing its asymptotic behavior \cite{Wen}.
Consider, for example, the equal time edge Green function, defined as
\begin{equation}
G_{edge}(x)={ <\Psi|\Psi_e^\dagger(x)\Psi_e(0)|\Psi>\over <\Psi|\Psi>}
\end{equation}
where $\Psi$ is the ground state, $\Psi_e$ and $\Psi_e^\dagger$ are annihilation and creation
field operators, and $x$ is the distance along the edge.
Wen argued that the long-distance behavior of the Green function is given by 
$|G_{edge}(x)|\sim x^{-\alpha}$ with $\alpha=3$ for $\nu=1/3$; such behavior was 
verified\cite{Lee} in an 
explicit calculation for Laughlin's wave function for bosons at $\nu=1/2$.

However, the exponent measured in the experiment of Grayson {\em et al.} \cite{Chang1}
is a continuous function of the filling factor.  More recent experiments 
\cite{Chang2} show evidence for a plateau near $\nu=1/3$ but with $\alpha\approx 2.7$.
This raises the question that we wish to explore in this work:  Does the actual form of the 
interaction affect the properties of the edge Luttinger liquid, and 
if so, how?  There has been much work investigating this and related issues 
\cite{Moon,Shytov,Zulicke,Goldman2}, and our goal here is 
to compute the exponent in a microscopic approach.

The microscopic understanding of the FQHE
is based on certain wave functions\cite{Laughlin,Jain}, 
which are known to be quite accurate for the bulk, homogeneous FQHE states from 
extensive comparisons against exact results in the edge-less spherical geometry.
However, the accuracy of these wave functions remains relatively untested at the edges.  
The validity of the wave functions in the bulk does not necessarily carry over to the edges. 
In the bulk, the ground state is separated from excitations by a gap, which renders it
robust to the actual form of the interaction.  On the other hand, at the edges, 
the system is compressible with gapless excitations, which may make it more susceptible to
various perturbations and the actual form of the inter-electron interaction.

We will work in the disk geometry, which has been employed by several authors in the past
\cite{Laughlin,Dev,Kawamura,Cappelli,Han,Tsiper}.  We neglect in the following mixing with 
higher electronic Landau levels, as appropriate in the large $B$ limit.
We also do not include in our
calculations any external confinement potential;  electrons are confined within a disk
because of the restriction on the total angular momentum.  (These results can be shown to be 
directly relevant to a weak parabolic confinement, which enters into the problem only 
through a renormalization of the magnetic length.)

We will focus on the filling factor 
at which the edge properties have been most investigated, namely the 
lowest Landau level (LL) filling $\nu=1/3$.  The bulk ground state here is well approximated 
by Laughlin's wave function \cite{Laughlin}:
\begin{equation}
\Psi^0_{1/3}=\prod_{j<k}(z_j-z_k)^{3} \exp\left[-\frac{1}{4 l_0^2}\sum_i |z_i|^2\right]\;,
\end{equation}
where $z_j=x_j-iy_j$ denotes the position of the $j$th particle as a complex number and
$l_0=\sqrt{\hbar c/eB}$ is the magnetic length.
In fact, $\Psi^0_{1/3}$ is the exact ground state for a short range model potential 
\cite{Haldane} denoted by $V_1$.  However, it is not exact for the Coulomb interaction, and 
significant deviations exist between this wave function and
the exact Coulomb ground state in the disk geometry \cite{Tsiper}.

\begin{figure}
\vspace{-2.0cm}
\centerline{\psfig{figure=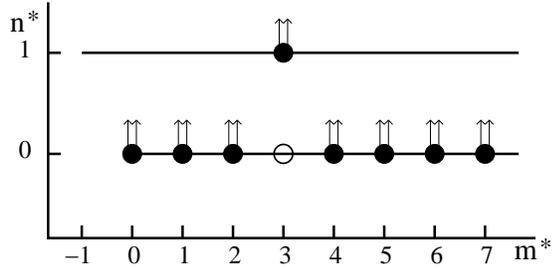,height=3.0in,angle=-90}}
\vspace{-1.5cm}
\caption{Schematic depiction of the composite-fermion exciton in which a composite fermion, 
shown as an electron bound to two flux quanta,  
is excited to a higher CF-Landau level.  This state is described by the 
wave function $\Psi_{m^*}^{CF-exciton}$, explained in the text, with $m^*=3$.
The y-axis is the composite-fermion Landau level index ($n^*$), and the x-axis is 
the effective angular momentum $m^{*}$.}
\label{fig1}
\end{figure}

To gain an understanding of the physical origin of the corrections to $\Psi^0_{1/3}$,
we consider the question of 
mixing of $\Psi^0_{1/3}$ with low energy excitations, as might be 
induced by the long range part of the Coulomb
interaction.  We construct excitations using the composite fermion theory of the 
FQHE \cite{Jain}.  In this approach, $\Psi^0_{1/3}$ is interpreted 
as one filled composite-fermion LL, rewritten it as 
\begin{equation}
\Psi^0_{1/3}=\Phi_1\prod_{j<k}(z_j-z_k)^{2} \;,
\end{equation}
where $\Phi_1$ is the wave function of one filled LL, given by 
\begin{eqnarray}
\Phi_{1}&=&
A[\eta_0(\vec{r}_1)\eta_1(\vec{r}_2)...\eta_{N-1}(\vec{r}_N)]\nonumber \\
&=&\prod_{j<k}(z_{j}-z_{k})\;
\exp[-\frac{1}{4l_0^2}\sum_{i}|z_{i}|^{2}]\;\;,
\end{eqnarray}
where $A$ is the antisymmetrization operator and $\eta_m(\vec{r})$ is the single particle wave
function in the lowest LL with angular momentum $m$.
The Jastrow factor $\prod_{j<k}(z_j-z_k)^{2}$ attaches to each electron in $\Phi_1$ 
to convert it into a composite fermion.   The excitations of $\Psi^0_{1/3}$ 
are then images of excitations in $\Phi_1$.  
In particular, a particle hole pair excitation in $\Phi_1$ produces a particle-hole pair 
of composite fermions, namely a composite-fermion (CF) exciton, in the full wave function. 
We restrict our attention below to the states containing a single CF exciton \cite{footnote}, 
and furthermore, to those excitons for which the CF-LL index changes by unity.
(Remember: mixing with higher {\em electronic} LLs is not considered in this work.)
The validity of this approximation will be checked below.
These excitons are related to excitations in which one electron in $\Phi_1$ is promoted from 
the angular momentum $m^*$ state in the lowest electronic LL 
to the angular momentum $m^{*}$ state in the second electronic LL, producing an excited state 
denoted by 
$$ \Phi_{1,m^*}^{el-exciton} = \hspace{2cm}$$ 
\begin{equation}
A[\eta_0(\vec{r}_1)...\eta_{m^*-1}(\vec{r}_{m^*})
\nonumber \\
\zeta_{m^*}(\vec{r}_{m^*+1})\eta_{m^*+1}(\vec{r}_{m^*+2})...\eta_{N-1}(\vec{r}_N)]\;
\end{equation}
where $\zeta_{m^{*}}$ is the wave function of an electron in the {\em second} LL in 
angular momentum $m^{*}$ state.
The excitations that do not conserve $m^*$  
are not relevant because they have a different total angular momentum than $\Psi^0_{1/3}$ and 
therefore do not couple to it. 
The corresponding CF exciton at $\nu=1/3$ is shown schematically in
Fig.~(\ref{fig1}), with the wave function given by:
\begin{equation}
\Psi_{m^*}^{CF-exciton}={\cal P}_{LLL} \Phi_{1,m^*}^{el-exciton} \prod_{j<k}(z_j-z_k)^{2} \;,
\end{equation}
where ${\cal P}_{LLL}$ denotes projection of the wave function into the lowest LL.  The 
explicit form of the projected wave functions can be written in the standard manner\cite{JK}.
The largest value of $m^{*}$, corresponding to the angular momentum of the outermost 
occupied state in $\Phi_1$, is $N-1$, giving a total of $N$ CF-exciton states.

\begin{minipage}{80mm}
\begin{table}[t]
\caption{The overlap ${<\Psi^0_{1/3}|\Psi_{1/3}^{mod}>\over \sqrt{<\Psi^0_{1/3}|\Psi^0_{1/3}>
<\Psi_{1/3}^{mod}|\Psi_{1/3}^{mod}>}}$ 
as a function of $N$, the number of particles.
The states are explained in the text.
Also given are $E^0=<\Psi^0_{1/3}|H|\Psi^0_{1/3}>/<\Psi^0_{1/3}|\Psi^0_{1/3}>$ and 
$E^{mod}=<\Psi_{1/3}^{mod}|H|\Psi_{1/3}^{mod}>/<\Psi_{1/3}^{mod}|\Psi_{1/3}^{mod}>$ in 
units of $e^2/\epsilon l_0$, where $H$ is the Coulomb interaction.  The 
quoted energy is the total energy of the interacting electron system (without taking into 
account any neutralizing background). 
\label{tab:Tab1}}
\vspace{0.4cm}
\begin{center}
\begin{tabular}{|c|c|c|c|c|c|}
$N$  & 10 & 11 & 12 & 20 & 30 \\ \hline
overlap & 0.982 & 0.980 & 0.979 & 0.971 & 0.965 \\ 
\hline
$E^0$ & 7.175 &  8.467 &  9.839 &  23.278 &  45.251 \\ \hline
$E^{mod}$ &  7.172 &  8.463 &  9.835 &  23.272 &  45.243 \\
\end{tabular}
\end{center}
\end{table}
\end{minipage}

\begin{figure}
\vspace{-0.7cm}
\centerline{\psfig{figure=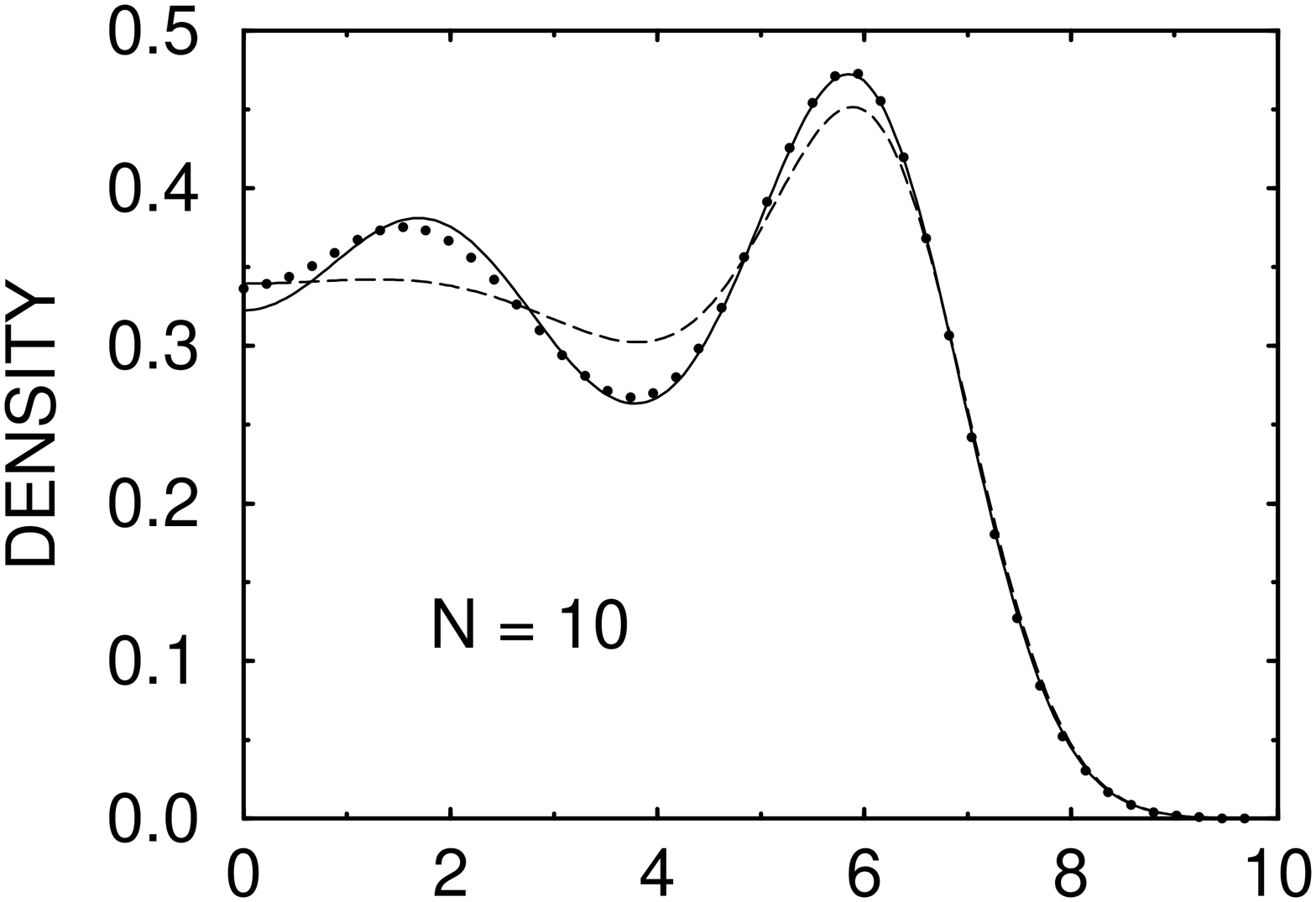,height=2.5in,angle=-90}}
\vspace{-1.5cm}
\centerline{\psfig{figure=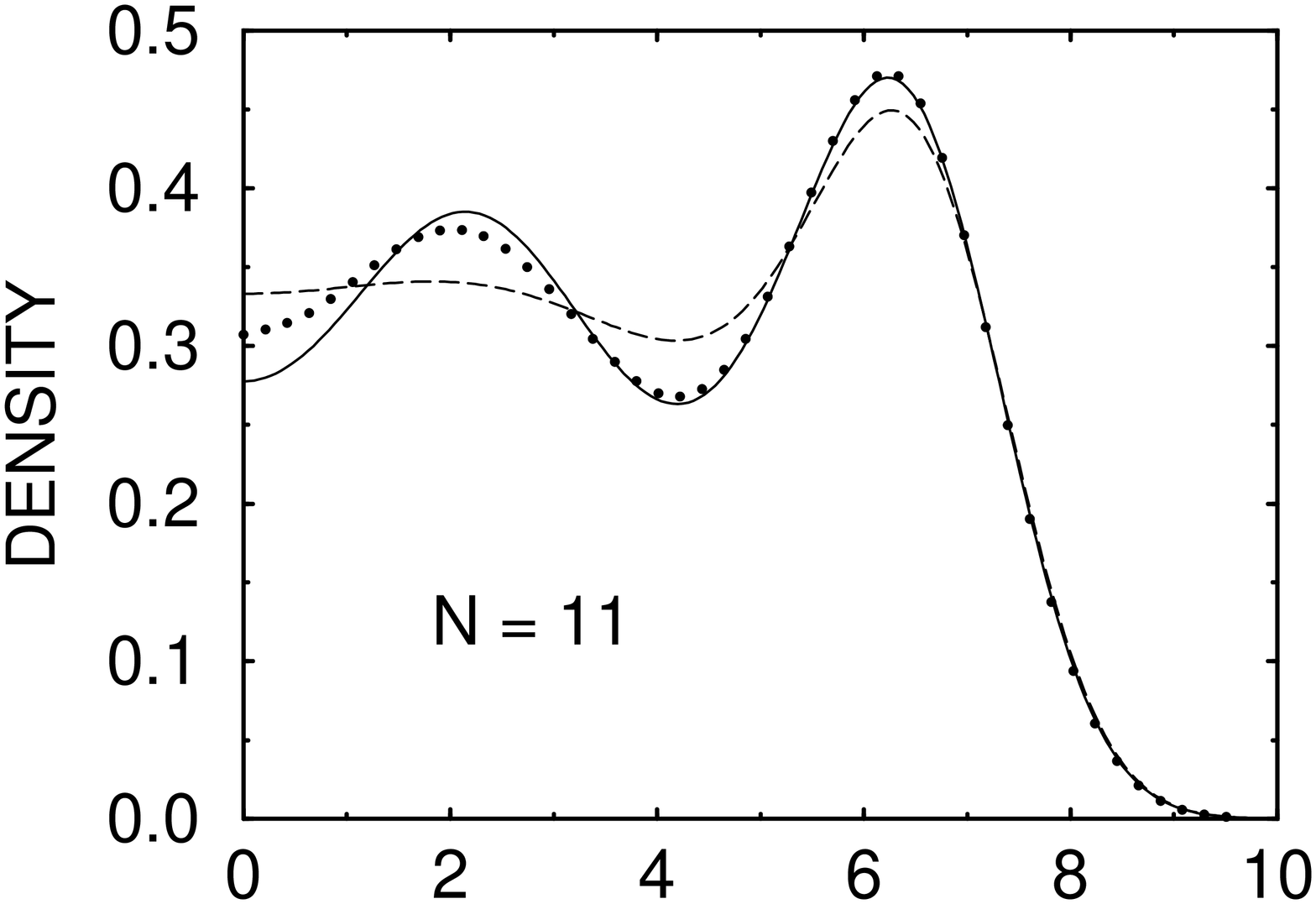,height=2.5in,angle=-90}}
\vspace{-1.5cm}
\centerline{\psfig{figure=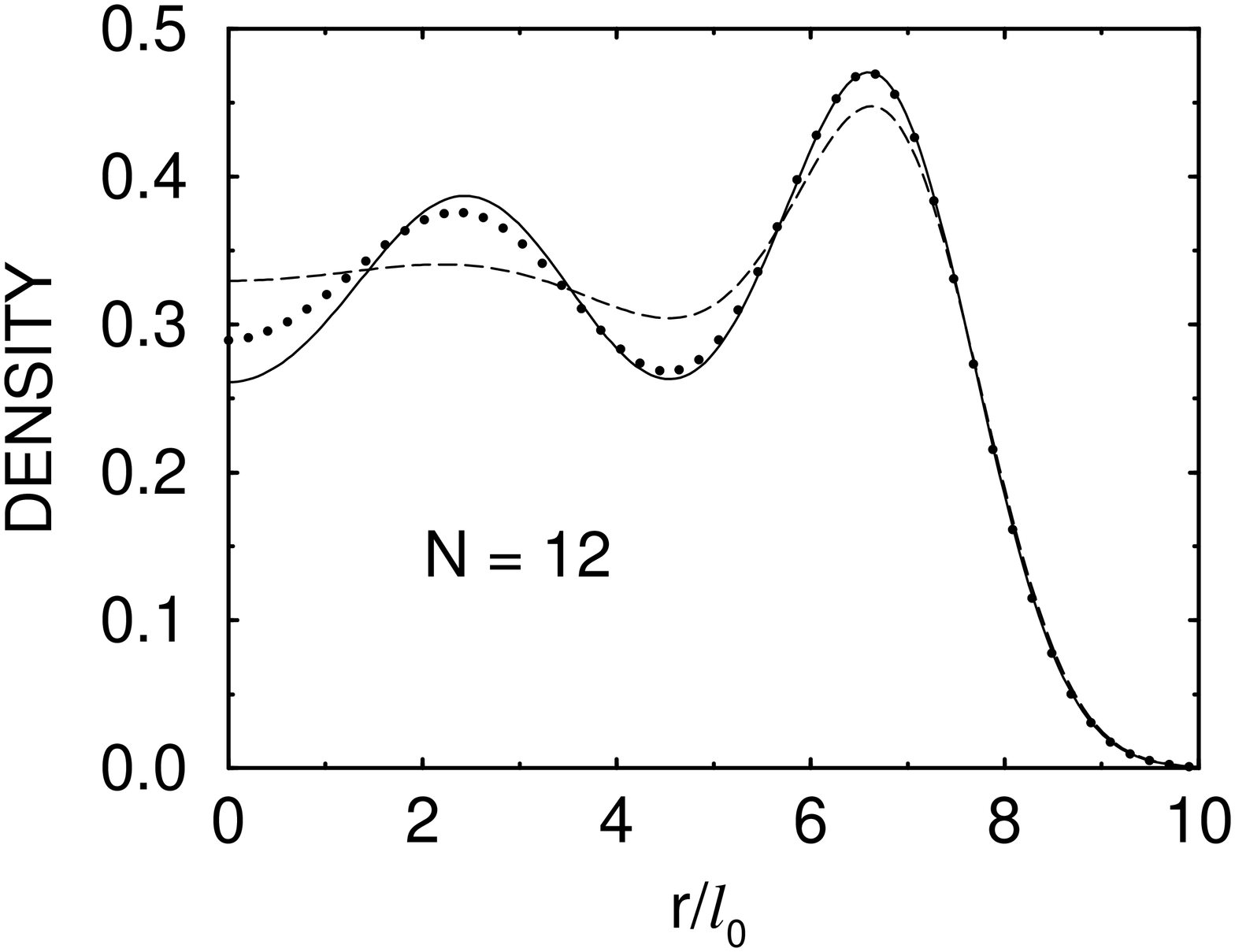,height=2.5in,angle=-90}}
\caption{The density profile of 
$\Psi^0_{1/3}$ (dashed line) and the modified composite fermion state
$\Psi_{1/3}^{mod}$ described in text (dots) for $N=10$, 11, and 12 particles.  
The distance from the center of the disk, $r$,  
is quoted in units of the magnetic length, $l_0=\sqrt{\hbar c/eB}$.  
The statistical uncertainty in Monte Carlo is smaller than the size of the dots. Also 
given is the density profile for the exact Coulomb ground state (solid line), 
taken from Tsiper and Goldman\protect\onlinecite{Tsiper}.} 
\label{fig2}
\end{figure}

\begin{figure}
\centerline{\psfig{figure=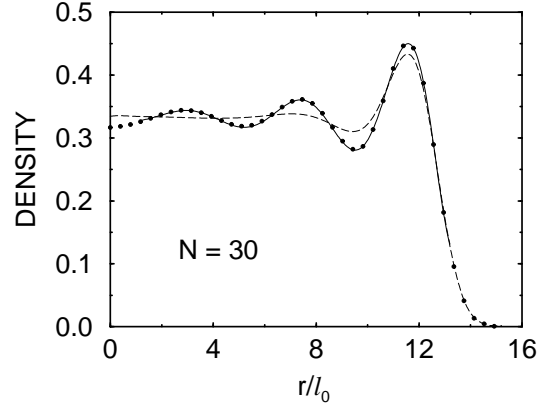,height=2.5in,angle=-90}}
\caption{The density profiles for $\Psi_{1/3}^0$ (dashed line) and $\Psi_{1/3}^{mod}$  
(dots) for $N=30$.  The solid line is a fit according to $\rho(r)=1/3 + \frac{\rho_1}{(1-\beta
x+\gamma x^2)} \cos\left(2\pi
\frac{x}{x_0-\alpha x}\right)$ with 
$x=r-r_0$, $r_0=11.407$, $\rho_1=0.113$, $x_0=4.02$, $\alpha=0.029$, $\beta=0.415$, and $\gamma=0.086$.
(All distances are quoted in units of $l_0$.)}
\label{fig3}
\end{figure}

We next proceed to diagonalize the Coulomb Hamiltonian in the 
restricted Hilbert space of $N+1$ states: $\Psi^0_{1/3}$
and $\Psi_{m^*}^{CF-exciton}$ (with $m^*=0,...,N-1$).  One difficulty is that these states
are not mutually orthogonal.
We obtain an orthogonal basis following the Gram-Schmid procedure, 
compute the matrix elements of the Coulomb Hamiltonian, and then carry out diagonalization 
in the restricted Hilbert space, performing all integrals by the Monte Carlo method.  
While the procedure is rather cumbersome, it can be carried 
out satisfactorily, and we refer the reader to literature for further details and also 
for a trick used to evaluate off-diagonal matrix elements by Monte Carlo \cite{Mandal}.
To obtain adequate accuracy, we
need to perform up to 10 million Monte Carlo iterations for each point.
The modified composite fermion ground state obtained in this manner, i.e., 
taking into account the mixing of
the ``unperturbed" CF ground state $\Psi^0_{1/3}$ with single CF 
excitons,  will be denoted by $\Psi_{1/3}^{mod}$.  We note that like $\Psi^0_{1/3}$, 
$\Psi_{1/3}^{mod}$ also involves only single particle states with $m\leq
3(N-1)$; the states with non-zero occupation of $m>3(N-1)$
would contain two or more CF-excitons, which we neglect in this work.  
We have studied several systems up to $N=30$; 
the overlap between $\Psi_{1/3}^{mod}$ and $\Psi^0_{1/3}$, also evaluated by Monte Carlo,
is shown in Table I.  The density profile of $\Psi_{1/3}^{mod}$ is shown in 
Figs.~(\ref{fig2}) and (\ref{fig3}) for several values of $N$.

To test the validity of the approximations made above, we compare in Fig~(\ref{fig2}) 
the modified density profile calculated above with the exact Coulomb 
density profile taken from Ref.~\onlinecite{Tsiper}.  The comparison demonstrates   
that $\Psi_{1/3}^{mod}$ is a significant improvement over $\Psi^0_{1/3}$, 
and is actually quite close to the exact wave function, thus establishing that much 
of the discrepancy between $\Psi^0_{1/3}$ and the exact wave
function can be accounted for by incorporating mixing with the lowest energy single 
CF-excitons.  This, to our knowledge, is the first variational improvement of the FQHE wave 
function in the lowest LL space.

Having demonstrated that $\Psi_{1/3}^{mod}$ correctly captures the 
interaction dependent physics at the edge,
we proceed to calculate the edge Green function.
For the disk geometry, the equal time edge Green function is defined as
\begin{equation}
G_{edge}(\theta)={ <\Psi|\Psi_e^\dagger(R_\theta)\Psi_e(R)|\Psi>\over <\Psi|\Psi>}
\end{equation}
where $\Psi$ is the ground state describing a disk of electrons centered at the
origin, $z=R$ is the positions at the edge along the 
x-axis and $z\equiv R_\theta=R e^{i\theta}$ is the position 
of the edge at an angle $\theta$ relative to the x-axis. 
Following Lee and Wen \cite {Lee}, we take $R=\sqrt{6N}\;l_0$ at $\nu=1/3$.
The power law exponent $\alpha$ is defined by the equation appropriate for 
the geometry under consideration\cite{Wen}:  
\begin{equation}
|G_{edge}(\theta)|\sim \sin^{-\alpha}(\theta/2)
\end{equation}
The Green function can be expressed as 
\begin{equation}
G_{edge}(\theta)= N \frac{\int d^2z_2...d^2z_N \Psi^*(R_\theta,z_2,...,z_N)\Psi(R,z_2,...,z_N)}
{\int d^2z_1...d^2z_N \Psi^*(z_1,...,z_N)\Psi(z_1,...,z_N)}
\end{equation}
We write the numerator as $\ll\Psi^*_\theta \Psi\gg$, where $\ll \gg$ denotes integral
over coordinates $z_2... z_N$. Using  
$$\ll \Psi^*_\theta \Psi\gg=\ll \Re\Psi^*_\theta \Psi +i \Im \Psi^*_\theta \Psi\gg $$
$$ 2 \ll \Re\Psi^*_\theta \Psi\gg =\ll |\Psi_\theta+\Psi|^2 \gg- \ll |\Psi|^2 \gg- 
\ll |\Psi_\theta|^2\gg $$
$$ 2\ll \Im\Psi^*_\theta \Psi\gg =\ll 
|\Psi_\theta+i\Psi|^2\gg -\ll |\Psi|^2\gg -\ll |\Psi_\theta|^2\gg $$
we express the real and imaginary parts of the edge Green function 
in terms of integrals over positive definite, real 
integrands, which can be evaluated efficiently by Monte Carlo.

\begin{figure}
\centerline{\psfig{figure=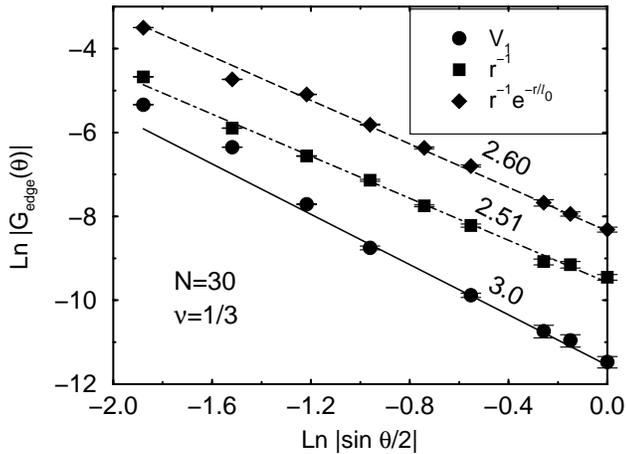,height=3.0in,angle=-90}}
\caption{The edge Geeen function as a function of the distance along the edge, parametrized by the
angle $\theta$.  The distance from the center is taken to be $R=\sqrt{6N}\;l_0$, $l_0$ being the
magnetic length.  The result is shown for $\Psi^0_{1/3}$, which is exact for the $V_1$
interaction, and the modified CF wave functions for two interactions: the Coulomb interaction
($r^{-1}$) and the Yukawa interaction ($r^{-1}e^{-r/l_0}$). In each case, the 
straight line is a fit with the exponent $\alpha$ shown near the line.}
\label{fig4}
\end{figure}

Fig.~(\ref{fig4}) shows $|G_{edge}(\theta)|$ for both $\Psi^0_{1/3}$ and 
$\Psi_{1/3}^{mod}$ for $N=30$. 
For the former wave function, we recover $\alpha=3$, as expected. The deviation of the Green function
from the power-law behavior
at short distances (small $\theta$) seen in Fig.~(\ref{fig4}) is quite similar to that found in 
Ref.~\onlinecite{Lee} for bosons at $\nu=1/2$. 
There is also a well defined exponent for $\Psi_{1/3}^{mod}$, 
indicating that the edge continues to be described as a Luttinger liquid.
However, the Luttinger exponent now is $\alpha \approx 2.5$.

The change in the exponent is surprisingly large considering the  
smallness of the corrections to $\Psi^0_{1/3}$.
Several points are in order here.  It is sometimes
stated that the arguments leading to $\alpha=3$ require a sharp edge, i.e., when 
the density is approximately constant (at its bulk value) up to the edge and falls sufficiently
sharply to zero at the edge.  Independent of the question of whether and how 
a sharp edge may be realized, we note that both
$\Psi^0_{1/3}$ and $\Psi_{1/3}^{mod}$ have rather similar density profiles at the edge,  
indicating  that sharp-vs-smooth-edge type considerations are not likely to have any bearing 
on the difference in the exponents.  
We have also tested that the mixing with the CF excitons does not affect the compressibility of 
the state; the gap in the angular momentum $L=3(N-1)$ sector of the Hilbert space 
is only very weakly affected by it.
In other words, the mixing with the second CF-LL merely renormalizes
the states of the lowest CF-LL, while preserving the single channel character of the problem.
Finally, while it is impossible to prove rigorously that the conclusion based on our finite 
system is valid in the thermodynamic limit, 
the system size ($N=30$) is sufficiently large to  
reproduce the expected exponent $\alpha=3$ for Laughlin's wave function and also to give a well
defined exponent for the modified wave function.

One might wonder if a universal exponent is obtained at least for all 
{\em short-ranged} interactions.  To address this issue, we have  also considered a 
short-ranged, Yukawa-type interaction $V(r)=r^{-1}\exp[-r/l_0]$.
The edge Green function is computed as before, also shown in Fig.~(\ref{fig4}), from which 
the edge exponent $\alpha\approx 2.6$ is deduced.   
This suggests that the exponent $\alpha=3$ is unique to 
$\Psi^0_{1/3}$, which has the very special property that all of the zeros of the wave
function are bound to the electrons; any splitting of the zeros away from electrons seems to 
change the exponent.

We believe that our study provides compelling evidence that the 
edge exponent is sensitive to the detailed form of the wave function.  
Because the Hall quantization is not affected by small perturbations in the wave 
function, so long as the gap in the bulk is preserved, one must conclude that the 
Luttinger parameter $\alpha$ is not tied to the 
quantized Hall conductance, and is therefore not a fundamental property of the FQHE state.
Investigations into the origin of the non-universality of the edge properties 
are bound to produce interesting physics.

Our results are generally consistent with the tunneling experiments of Chang and 
collaborators \cite{Chang1,Chang2}.  The theoretical 
exponent obtained above is smaller than 3.0 for both the
Coulomb and Yukawa interactions, consistent with experiment. (A more accurate
estimation of the observed exponent will require a realistic treatment 
of the confining potential at the edge\cite{Zulicke}.)  
Also, unless the exponent is completely fixed by the Hall 
conductance, there is no reason to expect it to be precisely constant 
over the range of a Hall plateau.

This work was supported in part by the National Science
Foundation under Grant No. DMR-9986806.  We thank Prof. V. Goldman for sharing his exact
diagonalization results with us, and thank S. Das Sarma, C. Kane, A.H. MacDonald, and 
G. Murthy for discussions.


\begin{references}
\bibitem{footnote}  We note that in the spherical geometry, the ground state does not mix with
the single CF-exciton (involving excitations to arbitrary CF-LL)
for symmetry reasons, because the CF excitons have non-zero orbital angular momenta. 
This indicates that the corrections to the FQHE wave function in the bulk involve mixing with 
higher order excitations, containing two or more CF-excitons, and hence are much weaker.
\end{references}
\end{document}